 \documentstyle[preprint,aps]{revtex}
 
\begin{document}
 \newcommand{\Qed}{\rule{2.5mm}{3mm}}
 \newcommand{\balpha}{\mbox{\boldmath {$\alpha$}}}
 \draft
 \title{Linear equations of motion for massless particles of any spin in any
 even dimensional spaces}
 \author{  Bojan Gornik}
 \author{ N. Manko\v c Bor\v stnik}
 \address{ Faculty of Mathematics and Physics, University of
 Ljubljana, Jadranska 19, \\ 
 Ljubljana 1000, Slovenia  \\
 and Primorska Institute for Natural Sciences and Technology,\\ 
 C. Mare\v zganskega upora 2, Koper 6000, Slovenia}
 \date{\today}
\maketitle
 
\begin{abstract}
It is proven that the Poincar\' e symmetry determines equations 
of motion, which are  for massless particles of any spin
in $d$-dimensional spaces  linear in the momentum: 
$(W^a=\alpha p^a)|\Phi\rangle$
with $W^a$ the generalized Pauli-Ljubanski vector.
The proof is made only for even $d$ and  for
fields with no gauge symmetry. We comment on a few examples. 
\end{abstract}

\section{Introduction.} \label{intro}

The ordinary space-time which we seem to experience is four-dimensional. However, besides the
ordinary space-time, the internal space of spin and charges is equally important. Without 
spins and charges the World would manifest no dynamics - noninteracting chargeless scalar fields 
could form no matter.
Theories of strings and membranes\cite{stringsmembranes} as well as 
Kaluza-Klein theories\cite{kaluzaklein}
are predicting more than four-dimensional ordinary space-time. One of 
us\cite{mankoc92,mankoc93,mankoc94,mankoc95,mankoc99,mankocborstnik98,mankocnielsen99,nf}
has proposed the approach,
which describes geometrically not only the ordinary space-time but also the internal space of spins and 
charges, unifying spins and charges.  In this approach the internal space is described by
a vector space spanned over the anticommuting coordinate space of the same dimension $d$ as that of ordinary
space-time. 
To describe the physics of the Standard model both spaces, the one of commuting and the one of anticommuting
coordinates, have to be more than four-dimensional. In $d$-dimensional space-time only spin degrees of freedom
exists. It is the appropriate break of symmetry, which in four-dimensional subspace makes the spin 
manifesting itself
as the spin and the known charges. 
In all theories with $d > 4$ the question arises why  Nature has made a choice of  four-dimensional
subspace with one time and three space coordinates and with the particular choice of charges beside the
spin degree of freedom for either spinors or vectors.

Manko\v c and Nielsen\cite{mankocnielsen00,mankocnielsen01}
proved that in $d$-dimensional spaces, with even $d$, the spin degrees of freedom require $q$ time and $(d-q)$ space
dimensions, with $q$ which has to be even. Accordingly in four-dimensional space Nature could only make a choice of 
the Minkowski metric. This proof was made under the assumption that equations of motion are for massless
fields of any spin linear in the $d$-momentum $p^a,\; a = 0,1,2,3,5,\ldots,d$.
(In addition, also the Hermiticity of the equations of motion operator
as well as that this operator operates within an irreducible representation of the Lorentz group was 
required.)
Our experiences tell us that equations of motion of all known massless fields are linear in the four-momentum $p^a,
a = 0,1,2,3.$ We are refering\cite{gornikmankoc} 
to the Dirac equation of motion for massless spinor fields and the Maxwell or
Maxwell-like equations of motion for massless vectorial fields. One of us together with 
A. Bor\v stnik\cite{mankocborstnik98} has shown that the Weyl-like equations 
exist not only for spinors but also for vectors.

In this paper we present the proof that  equations of motion for free massless particles in even-dimensional 
spaces, if manifesting the Poincar\' e symmetry, are linear in the $p^a-$momentum. For four
dimensional space-time Wigner \cite{wigner} clarified that 
as far as  masses and  spins are concerned (point-like) particles can be described by their properties 
under transformations of the Poincar\' e group. The classification of  particles with respect to
the unitary discrete representations of the Poincar\' e group can be found in Weinberg\cite{weinberg2}, for example.
In his language  equations of motion are
those equations that constrain a solution space to a certain Poincar\' e group representation. 
For spinors this leads to the Dirac equation and for vectors to the Maxwell equations\cite{fonda}.
The aim of this paper is to use similar
techniques for massless particles in general even-dimensional spaces. We 
prove that in even-dimensional  spaces, for any $d=2n$, free massless fields $|\Phi\rangle$

\begin{equation}
\label{kg}
(p^a p_a = 0)\; |\Phi\rangle,\quad a= 0,1,2,3,5,\cdots,2n
\end{equation}
of any spin satisfy
equations of motion which are  linear in  a $d$-momentum $p^a = (p^0, \overrightarrow{p})$ 
\begin{equation}
(W^a = \alpha p^a)\;|\Phi\rangle,
\quad a = 0,1,2,3,5,\dots,d,
\label{meq}
\end{equation}
with $\alpha = \frac{\overrightarrow{S}\cdot \overrightarrow{p}}{|p^0|}$ to be determined in this 
paper\footnote[1]
{After this paper appeared on hep-th W. Siegel let us know that
equations $(S^{ab}p_b + w p^a=0)|\Phi\rangle$
are the linear equations in $p^a$-momentum, as well as in $S^{ab}$, 
which all irreducible representations of massless fields in any $d$ obey,
and that these equations can be found in his book\cite{wsb} and in his 
paper\cite{wsp}. Following derivations of this paper in section (\ref{littlegroup})
one easily proves that solutions of Siegel's equations belong to irreducible representations
of the Poincar\' e group. The proof is  simpler than for  our equations 
$(W^a=\alpha p^a)|\Phi\rangle$. Both equations are of course equivalent. Following our derivations
one finds that the constant $w$ in  the Siegel's equation is $w=l_n$, with $l_n$ defined in
Eq.(\ref{dejstvo2}).
One derives our equation from  the Siegel's one
for even $d$ after some  rather tedious calculations if multiplying it by 
$\varepsilon_{aca_1a_2 \ldots a_{d-3} a_{d-2}} S^{a_1 a_2} \cdots
S^{a_{d-3}a_{d-2}}$. The way how we present the linear equations of motion has 
several obvious advantages, which reader will easily find by himself.}.
For spinors, which will be determined in  Eq.(\ref{spinor}), $\alpha = \pm\frac{1}{2}$. 
The proof is made only for fields with no gauge 
symmetry and with a nonzero value of the {\it handedness} operator\cite{mankoc93} $\Gamma^{(int)}$
\begin{equation}
  \Gamma^{(int)} = \beta \,\varepsilon_{a_1 a_2 \ldots a_{d-1}a_d} S^{a_1 a_2} \ldots
   S^{a_{d-1}a_d},
  \label{gamaint}
\end{equation}
which commutes with all the generators of the Poincar\' e group.
We choose $\beta$  so that $\Gamma^{(int)} = \pm 1$. For spinors (Eq.(\ref{spinor}))  $\beta = i \frac{2^n}{(d!)}$,
while for  fields of a general spin $\beta $ will be
 determined in section \ref{representations}.  Operators $S^{ab}$ are the generators of the
 Lorentz group $SO(1, d-1)$ in internal space, which is the space
of spin degrees of freedom.
In Eq.(\ref{meq}) $W^a$ is the generalized Pauli-Ljubanski\cite{mankoc93} $d$-vector
\begin{equation}
  W^a = \rho \,\varepsilon^{ab}{}_{a_1 a_2\ldots a_{d-3}a_{d-2}} p_b S^{a_1 a_2}\ldots
    S^{a_{d-3}a_{d-2}}.
  \label{pauliljub}
\end{equation}
We define \cite{mankocnielsen01} $S^i$ as the $d-1$ vector
\begin{equation}
S^i = \rho  \,\varepsilon^{0i}{ }_{a_1a_2 \cdots a_{d-3} a_{d-2}}
 S^{a_1a_2} \cdots S^{a_{d-3}a_{d-2}}
\label{svector}
\end{equation}
and determine the value of $\rho$  so that the eigenvalues of $S^i$ are independent of a dimension and
the same as in four-dimensional space. We will show that this dictates the choice
$\rho = \frac{2^{n-2}}{(d-2)!}$ for spinors ($S^i = \pm\frac{1}{2}$) and 
$\rho = \frac{1}{2^{n-1}(n-1)!^2}$ for vectors ($S^i = \pm 1, 0$).
Eq.(\ref{meq}) also guarantees that Eq.(\ref{kg}) is fulfilled for all massless fields with no gauge freedom
and with nonzero spin.

We prove that for spinors in $d$-dimensional space Eq.(\ref{meq}) is equivalent to the equation
\begin{equation}
(\Gamma^{(int)} p^0 = \frac{1}{|\alpha|} \;\overrightarrow{S}\cdot \overrightarrow{p})|\Phi\rangle.
\label{meq1}
\end{equation}

with $\frac{1}{|\alpha|}$ equal to $2$ for any d=2n, while for a general spin Eq.(\ref{meq})  may
impose additional conditions on the field. 

We recognize the generators $S^{ab}$ to be of the spinorial character, if they fulfil the relation
\begin{equation}
\{S^{ab}, S^{ac}\} = \frac{1}{2} \; \eta^{aa}\; \eta^{bc}, \;\;{\rm no\;\; summation\;\; over\;\; a},
\label{spinor}
\end{equation}
with $\{A,B\} = AB + BA $.

In this paper the metric is, independently of the dimension,  assumed  for simplicity to be the 
Minkowski metric with 
$\eta^{ab}= \delta^{ab} (-1)^A, \;A= 0,\; {\rm for}\; a=0 \;{\rm and}\; A=1$, otherwise.

The paper is arranged as follows:

We first present in section \ref{poincare} the infinitesimal generators of the 
Poincar\'e group in $d$-dimensional 
spaces, the corresponding algebra and the Casimirs. We define appropriate $d$-vectors and 
$(d-1)$-vectors  (some of them can be defined only in even-dimensional
spaces),
and present their properties and  their 
commutation (and for spinors anticommutation) relations. 

In section \ref{representations} we review
representations of the Lorentz group and determine the parameter $\beta$ of Eq.(\ref{gamaint}), specifying the Casimir
$\Gamma^{(int)}$ for a general spin.

In  section \ref{littlegroup} we  define the generators of the little group and the constraints, which the
generators of the little group have to fulfil in order to define discrete representations of the
Poincar\' e group. 
We determine the factor $\alpha$ in Eq.(\ref{meq}),  so that the
equation Eq.(\ref{meq}) holds on discrete representations of the Poincar\' e group and 
 represents accordingly a promising candidate for  equations of motion.

In section \ref{equationsofmotions} we investigate the solutions of Eq.(\ref{meq})
on spaces with $\Gamma^{(int)}\not=0$. We 
show that they form irreducible representations of the Poincar\' e group (barring the sign of energy degeneracy)
 and in this sense we 
prove that Eq.(\ref{meq}) is the equation of motion.

At the end we comment in section \ref{discussions} on spinorial representations 
and on  vectorial (we shall define these representations later) representations 
in any $d=2n$-dimensional spaces. 

\section{Poincar\' e symmetry.} \label{poincare}

The generators of the Poincar\' e group, that is the generators of translations $p^a$ and 
the generators of the Lorentz transformations $M^{ab}$ (which form the Lorentz group),
fulfil in any dimension $d$, even or odd, the commutation relations:
\begin{eqnarray}
[p^a, p^b] &=& 0,
\nonumber
\\
\;[M^{ab}, M^{cd}]
&=&  i(\eta^{ad} \; M^{bc} + \eta^{bc} \; M^{ad} - \eta^{ac} \; M^{bd} - 
\eta^{bd}\; M^{ac}),
\nonumber
\\
\;[M^{ab}, p^c]
&=& i (\eta^{bc}\; p^a - \eta^{ac}\; p^b).
\label{poincarealgebra}
\end{eqnarray}
The generators of the Lorentz transformations of the internal group $S^{ab}$ ($M^{ab} = 
L^{ab} + S^{ab} $, with $L^{ab} = x^a p^b - x^b p^a $)  fulfil the same 
commutation relations as $M^{ab}$ (as do also $L^{ab} $ by themselves) and commute with $p^a$
(commutation relations for $[L^{ab},p^c]$ are the same as for $M^{ab}$). 

There are $n$ commuting operators of the Lorentz group in $d$-dimensional spaces,
for either $d=2n+1$ or $d=2n$, namely $M^{01}, M^{23}, \cdots, M^{d-1\; d}$ (or any other
set of generators with all different indices) and accordingly $n$ Casimirs of the 
Lorentz group. There are two Casimirs of the Lorentz group\cite{mankoc93,mankoc95}, which are
easily found to be:
\begin{eqnarray}
M^2 :&=& \frac{1}{2}\; M^{ab} M_{ab}\quad {\rm and} 
\nonumber
\\
\Gamma :&=& \beta
\; \varepsilon_{a_1 a_1  \cdots a_{2n-1} a_{2n}}
\; M^{a_1a_2} \cdots M^{a_{2n-1}a_{2n}}.
\label{mg}
\end{eqnarray}
The second Casimir of Eq.(\ref{mg}) can only  be defined  for $d=2n$.
We see that $\Gamma^{(int)}$ from Eq.(\ref{gamaint}) follows from $\Gamma$ by replacing $M^{ab}$ 
by $S^{ab}$ and that $\Gamma^{(int)}$ commutes with all the generators of the Poincar\' e group.  

With the help of $\Gamma$ the generalized Pauli-Ljubanski vector as presented in 
Eq.(\ref{pauliljub}) 
can be defined as
\begin{eqnarray}
W^a = -i\frac{\rho}{2n\beta} [\Gamma, p^a],
\label{pl1}
\end{eqnarray}
with $\beta $ and $\rho$ from Eqs.(\ref{gamaint}, \ref{pauliljub}) for any even $d$.
$W^a$ in Eq.(\ref{pl1}) can be as well expressed by $S^{ab}$ instead by $M^{ab}$.
One immediately finds that
\begin{eqnarray}
W^a p_a = 0.
\label{wp}
\end{eqnarray}

\vskip0.5cm
 {\it Lemma \ref{poincare}.1}:
$W^a W_a $ is the   Casimir of  the Poincar\' e group
and so is $p^a p_a$.

\vskip0.2cm

{\it Proof}: The proof goes, by taking into account Eq.(\ref{pl1}) 
and the Lie algebra of the Poincar\'e group, as follows
\begin{eqnarray}
[W^a W_a, M^{bc} ]  = -i \frac{\rho}{2n\beta}
\{ W_a \; [[\Gamma,p^a], M^{bc}] + [[\Gamma,p^a], M^{bc}]\; W_a \}=
\nonumber \\
= -i \frac{\rho}{2n\beta} \{ W_a \;[\Gamma,[p^a, M^{bc}]] + [\Gamma,[p^a, M^{bc}]]\; W_a \} = 0.
\label{cascom1}
\end{eqnarray}
The proof that $W^a W_a$ commutes with $p^b$ is obvious due to Eq. (\ref{pauliljub}), while the
proof that $p^a p_a $ commutes with $M^{bc}$ is straightforward. \hfill \Qed
\vskip0.5cm

The following commutation relations follow
\begin{eqnarray}
\;[p^a,W^b] &=& 0,
\nonumber\\
\;[M^{ab}, W^c]&=& -i (\eta^{ac} W^b - \eta^{bc} W^a),
\nonumber\\
\;[\frac{1}{p^0}, W^a] &=& \frac{i}{(p^0)^2}\; (\eta^{a0} p^b-\eta^{b0} p^a).
\label{wpmab}
\end{eqnarray}

To prove the first equation of Eqs.(\ref{wpmab}) is straightforward. To prove the second one, 
Eq.(\ref{pl1}) as well as Jacobi identity $[M^{ab}, [\Gamma, p^c]] + [p^c,[M^{ab},\Gamma]]
+ [\Gamma,[p^c, M^{ab}]] =0$
have to be taken into account. To prove the third of the above equations the equations 
$[M^{ab}, \frac{p^0}{p^0}] = 0 = \frac{1}{p^0}\; [M^{ab}, p^0] + [ M^{ab}, \frac{1}{p^0}] p^0$  have to be
taken into account. This type of a proof was used also in the ref.(\cite{fonda}).

We present, for the spinorial case only, the commutation and anticommutation relations for 
$(d-1)$-vectors $S^i$, which are defined in Eq.(\ref{svector})
\begin{equation}
[S^i, S^j] = i (-1)^{i-j} S^{ij},\quad \{S^i, S^j\} = \frac{1}{4} \eta^{ij}.
\label{spinorsi}
\end{equation}

\section{Representations of  Lorentz group in internal space.} \label{representations}

In this section we introduce the notation for ireducible representations of the Lorentz 
group $SO(1, d - 1)$ (or $SO(d)$) for $d=2n$, denoting an irreducible representation by
the weight of the dominant weight state of the representation. This exposition of the subject
follows those of \cite{giorgi}, \cite{matematika}. We also 
express the two Casimirs of Eq.(\ref{mg}) in terms of the dominant weight.
We pay attention only to the internal degrees of freedom for the Lorentz group, 
that is to a spin.

The Lie algebra of $SO(1, d-1)$ is spanned by the generators $S^{ab}$, which satisfy the commutation
relations of the second equation of Eqs.(\ref{poincarealgebra})
\begin{equation}
[S^{ab}, S^{cd}] = 
  i(\eta^{ad} \; S^{bc} + \eta^{bc} \; S^{ad} - \eta^{ac} \; S^{bd} - 
\eta^{bd}\; S^{ac}).
\label{komrel}
\end{equation}
  We choose  the $n$ commuting operators  of the Lorentz group $SO(1,d-1)$ as follows
  \begin{equation}
-i S^{0 d}, S^{1 2}, S^{3 5}, \ldots, S^{d-2 \; d-1}
\label{cartan}
\end{equation}
and call them
${\cal C}_0, {\cal C}_1, {\cal C}_2, \ldots, {\cal C}_{n-1}$, respectively.

(Everything what follows will be 
valid also for the group $SO(d)$, provided that the generators $-i S^{01}, -i S^{02}, \ldots$ 
are corespondingly replaced, that is by
$S^{01}, S^{02}, \ldots$, respectively.)

We say that a state $|\Phi_w\rangle$ has the weight
$(w_0, w_1, w_2, \ldots, w_{n-1})$ if the following equations hold
\begin{equation}
 {\cal C}_j|\Phi_w\rangle = w_j|\Phi_w\rangle,\quad j=0,1,\ldots,n-1. 
\label{weight}
\end{equation}

According to the definition in Eq.(\ref{cartan}), weight components
$w_0, w_1, w_2, \ldots, w_{n-1}$ are always real numbers.

  We introduce the raising and lowering operators
\begin{equation}
 E_{j k}(\lambda, \mu) := \frac{1}{2}((-i)^{\delta_{j 0}} S^{j_- k_-} + i\lambda S^{j_+ k_-}
 - i^{1+\delta_{j 0}}\mu S^{j_- k_+} - \lambda \mu S^{j_+ k_+} ), 
\label{raislow}
\end{equation}
with $0\le j<k\le n-1,\,\, \lambda,\mu=\pm 1$ and $0_- = 0, 0_+ = d, 1_- = 1, 1_+ = 2,
2_- = 3, 2_+ = 5$ and so on. For these operators the following commutation relations hold
\begin{equation}
  [ E_{j k}(\lambda, \mu), {\cal C}_l ] = (\delta_{j l}\lambda + \delta_{k l}\mu)  E_{j k}(\lambda, \mu). 
\label{raisinglowering}
\end{equation}
Therefore, if the state 
$|\Phi_w\rangle$ has the weight $(w_0, w_1, \ldots, w_{n-1})$ then the state 
$E_{j k}(\lambda, \mu) |\Phi_w\rangle$ has the weight $(\ldots, w_j + \lambda, \ldots, w_k + \mu, \ldots)$.

We now proceed to the definition of the dominant weight. We fix
$q\in\lbrace +1, -1\rbrace$ and call the state $|\Phi_{l}\rangle$ with the property
\begin{eqnarray}
  E_{j k}(q, \pm 1)|\Phi_{l}\rangle = 0, \quad 0\le j < k \le n - 1
\label{domweight}
\end{eqnarray}
 the dominant weight state. This state is (up to a scalar multiple) uniquely determined by  the
irreducible representation in question and vice versa, knowing the dominant weight state of a 
particular representation (or any state of a particular representation), all the others are
obtained by the application of the generators $S^{ab}$. We will denote the irreducible 
representation of the Lorentz group
$SO(1, d-1)$ by the weight of the dominant weight state: 
$(l_n, l_{n-1}, \ldots, l_2, l_1)_q$ with the index $q$ attached to distinguish which definition of the
dominant weight state we are using.
Numbers $l_n, l_{n-1}, \ldots, l_2, l_1$ are either all integer or all half integer and 
satisfy
\begin{equation}
  l_n\ge l_{n-1}\ge \ldots \ge l_2\ge |l_1|
  \label{dejstvo}
\end{equation}
in case $q=+1$ or
\begin{equation}
  l_n\le l_{n-1}\le \ldots \le l_2\le -|l_1|
  \label{dejstvo2}
\end{equation}
in case $q=-1$.

  The correspondence between the two notations is given by the fact that the following
representations
are
equivalent
\begin{equation}
(l_n, l_{n-1}, \ldots, l_2, l_1)_{+1}=(-l_n, -l_{n-1}, \ldots, -l_2, (-1)^{n-1} l_1)_{-1}.
\label{equiv}
\end{equation}
In section \ref{equationsofmotions} we will find it useful to work with the notation
$q=+1$ when dealing with the positive energy ($p^0 > 0$) representations and with
$q=-1$ when dealing with the negative energy ($p^0 < 0$) representations.

  For future use we reformulate the condition (\ref{domweight}) which determines the dominant
weight state. It follows from Eq.(\ref{raislow}) that
\begin{eqnarray}
 -i(E_{0 k}(q, +1) + E_{0 k}(q, -1)) &=& -S^{0 k_-} + q S^{d k_-}, 
 \nonumber
 \\
 - E_{0 k}(q, +1) + E_{0 k}(q, -1) &=& -S^{0 k_+} + q S^{d k_+} 
\label{dom1}
\end{eqnarray}
for $k=1, 2, \ldots, n - 1$. Eq.(\ref{domweight}) then implies
\begin{equation}
(S^{0 i} + q S^{i d}) |\Phi_{l}\rangle = 0, \quad i = 1, 2, 3, 5, \ldots, d - 1 . 
\label{dom2}
\end{equation}
Similarly, we obtain the following conditions
\begin{eqnarray}
(S^{1 i} + q i S^{2 i}) |\Phi_{l}\rangle = 0, \quad i = 3, 5, \ldots, d - 1,
\label{dom3} \\
(S^{3 i} + q i S^{5 i}) |\Phi_{l}\rangle = 0, \quad i = 6, 7 \ldots, d - 1,\,\,\,{\rm and\,\,so\,\,on.}
\label{dom4}
\end{eqnarray}
 By reversing this process we can also conclude that Eqs.(\ref{dom2}),
(\ref{dom3}),(\ref{dom4}) imply (\ref{domweight}). They are therefore equivalent to
condition (\ref{domweight}).

  We now determine the values of the two Casimirs of Eq.(\ref{mg}) for a
particular irreducible representation 
  $(l_{n}, l_{n-1}, \ldots, l_2, l_1)_q$.
  Since the Casimirs (Eq.(\ref{mg})) are scalars  it suffices to determine
their value on the dominant weight state $|\Phi_{l}\rangle$. Since the dominant weight state
satisfies Eqs.(\ref{dom2}),(\ref{dom3}),(\ref{dom4}) which are not invariant to the
permutations of ${\cal C}_0, {\cal C}_1, \ldots, {\cal C}_n$, we don't expect the values of
Casimirs to be symmetric in $l_n, l_{n-1}, \ldots, l_1$. It will turn out this is indeed the
case.

  We now find for the Casimir $M^2$ (which we shall denote when useful by $M^2_d$) from Eq.(\ref{mg})
\begin{equation}
  M^2_d = \frac{1}{2} S^{ab} S_{ab} = -(S^{0d})^2 - S^{0 i} S^{0 i} + S^{d i} S^{d i} + 
    \frac{1}{2}S^{i j} S_{i j}.
\label{m2d}
\end{equation}
We use Eq.(\ref{dom2})  and the relation $ S^{0 i} S^{0 i} - S^{d i} S^{d i}= 
(S^{0 i}  - q S^{d i})( S^{0 i} + q S^{d i}) -
q [ S^{0 i} , S^{d i}] $ to find
\begin{eqnarray}
 (S^{0 i} S^{0 i} - S^{d i} S^{d i})|\Phi_{l}\rangle &=& iq(d-2) S^{0d}|\Phi_l\rangle, \quad \Longrightarrow
\nonumber\\
M^2_d|\Phi_{l}\rangle &=& (-q(d-2)i S^{0d}-(S^{0d})^2 + \frac{1}{2} S^{i j} S_{i j} )|\Phi_{l}\rangle,
\nonumber\\
M^2_d|\Phi_{l}\rangle &=& ((d-2)q l_n + l_n^2 +   M^2_{d-2}) |\Phi_{l}\rangle.
\label{indukcijam2}
\end{eqnarray}
The operators $S^{i j}$ belong to the Lie algebra of the subgroup $SO(d-2)$ of the group $ SO(1, d-1)$ acting
only on coordinates $1, 2, \ldots, d-1$. We may use  Eq.(\ref{indukcijam2}) inductively to express the 
Casimir $M^2_d$ 
\begin{equation}
  M^2_d |\Phi_l\rangle= (l_n(l_n+q(d-2)) + l_{n-1}(l_{n-1}+q(d-4)) + \dots + l_2(l_2 + 2 q) + l_1^2)|\Phi_l\rangle. 
\label{m2r1}
\end{equation}

 We proceed with the Casimir $\Gamma^{(int)}$ of Eq.(\ref{gamaint}) 
in a similar way as with $M^2$ (again sometimes using the subscript $d$ to point out the dimension of space-time). 
We define
\begin{equation}
 \tilde{\Gamma}^{(int)}_d := \Gamma^{(int)}/ \beta =  \varepsilon_{a_1 a_2 \ldots a_{d-1}a_d} S^{a_1 a_2} S^{a_3 a_4} 
 \ldots \nonumber
\end{equation}
and note
\begin{eqnarray}
  \tilde{\Gamma}^{(int)}_d/(2n) = S^{0d}\varepsilon_{0d a_1 a_2 \ldots a_{d-3}a_{d-2}} S^{a_1 a_2} S^{a_3 a_4} \ldots + \nonumber \\
 +  2(n-1)\varepsilon_{0i dj a_1 a_2 \ldots a_{d-5}a_{d-4}} S^{0i} S^{dj} S^{a_1 a_2}S^{a_3 a_4}\ldots.
  \label{gammatilde}
  \end{eqnarray}

Applying $  \tilde{\Gamma}^{(int)}_d/(2n)$ on  the dominant weight state and taking into account
Eq.(\ref{dom2}) we find
\begin{eqnarray}
(\tilde{\Gamma}^{(int)}_d/(2n) - S^{0d}(\varepsilon_{0d a_1 a_2 \ldots a_{d-3}a_{d-2}} S^{a_1 a_2} S^{a_3 a_4} \ldots)
|\Phi_{l}\rangle = 
\nonumber\\
=(n-1) (\varepsilon_{0idj a_1 a_2 \ldots a_{d-5}a_{d-4}} (S^{0i}S^{dj}- S^{0j}S^{di} )S^{a_1 a_2}S^{a_3 a_4}\ldots)|\Phi_l\rangle=
  \nonumber\\
=(n-1)q (\varepsilon_{0idj a_1 a_2 \ldots a_{d-5}a_{d-4}} (S^{0i}S^{0j}- S^{0j}S^{0i} )S^{a_1 a_2}S^{a_3 a_4}\ldots)|\Phi_l\rangle=
  \nonumber\\
=(n-1)q (\varepsilon_{0idj a_1 a_2 \ldots a_{d-5}a_{d-4}} [S^{0i},S^{0j}]S^{a_1 a_2}S^{a_3 a_4}\ldots)|\Phi_l\rangle=
  \nonumber\\
=i(n-1)q (\varepsilon_{0dij a_1 a_2 \ldots a_{d-5}a_{d-4}} S^{ij}S^{a_1 a_2}S^{a_3 a_4}\ldots)|\Phi_l\rangle.
  \nonumber
\end{eqnarray}

It therefore follows that 
\begin{eqnarray}
  \tilde{\Gamma}^{(int)}_d|\Phi_{l}\rangle = 2n i(l_n + q(n - 1)) (\varepsilon_{0d a_1 a_2\ldots a_{d-5}a_{d-4}} S^{a_1 a_2}
  S^{a_3 a_4}\ldots)  |\Phi_{l}\rangle = 
  \nonumber\\
   2n i ( l_n + q(n - 1)) \; \tilde{\Gamma}^{(int)}_{d-2} |\Phi_{l}\rangle,
\label{indukcijagamma1}
\end{eqnarray}
where we have taken into account that $\Gamma^{(int)}_{d-2}$ refers to the handedness operator
for subgroup $SO(d-2)\le SO(1, d-1)$ acting on coordinates $1, 2, \ldots, d-1$.
We could repeat this process for $\Gamma^{(int)}_{d-2}$ with equations
\begin{equation}
(S^{1 i} +  i q S^{2 i}) |\Phi_{l}\rangle = 0, \quad i = 3, 5, \ldots, d - 1 
\nonumber
\end{equation}
taking place of  Eqs.(\ref{dom2}).  We obtain
\begin{eqnarray}
  \tilde{\Gamma}^{(int)}_{d-2}|\Phi_{l}\rangle = 2(n-1) (l_{n-1} + q(n - 2)) 
  (\varepsilon_{0d 12 a_1 a_2\ldots a_{d-3}a_{d-2}} S^{a_1 a_2}S^{a_3 a_4}\ldots)|\Phi_{l}\rangle = 
  \nonumber\\
  = 2(n-1) ( l_{n-1} + q(n - 2)) \; \tilde{\Gamma}^{(int)}_{d-4} |\Phi_{l}\rangle.
\nonumber
\end{eqnarray}
We note the absence of a factor $i$ in the last equation. Repeating this process inductively
we find
\begin{equation}
 ( \tilde{\Gamma}^{(int)}_d   = 2^n n! i (l_n + q(n - 1))(l_{n-1} + q(n - 2))\ldots(l_2 + q) l_1)|\Phi_l\rangle. 
\label{gama1}
\end{equation}
(Had we been dealing with the group $SO(d)$ the result would have been similar, the only 
difference being the absence of $i$  in the previous equation.)
We have therefore obtained
\begin{eqnarray}
\Gamma^{(int)} \; |\Phi_{l}\rangle= 2^n n! i\beta \; \prod_{j=1}^n \;\;
(l_j + q(j - 1))|\Phi_l\rangle.
\label{gamaf}
\end{eqnarray}

It follows from (\ref{gama1}) that on spaces $(l_n,\ldots,l_1)_{+1}$
with nonzero handedness
($\tilde{\Gamma}^{(int)}_d\not=0\Leftrightarrow l_1\not=0$) we must take
\begin{equation}
  \beta = \frac{i}{2^n n! (l_n+n-1)\ldots(l_2+1)|l_1|}, \quad d=2n,
  \label{beta}
\end{equation}
to obtain $\Gamma^{(int)} = \pm 1$. In what follows we assume this choice of $\beta$ has been made.

\section{Little group.} \label{littlegroup}

  In this section we characterize the unitary massless discrete representations of the Poincar\' e
group and work out some of their properties. In doing so we follow the little group method 
\cite{weinberg2}.

  For momenta $p^a$ appearing in an irreducible massless representations of the Poincar\' e group 
it holds 
\begin{equation}
p^a p_a = 0,\quad  p^0 > 0 \,\,{\rm or}\,\, p^0 < 0
\label{prvic}
\end{equation}
(we omit the trivial case $p^0=0$). We denote $r=\frac{p^0}{|p^0|}$. 
The Poincar\' e group representation is then characterized by the representation of the 
so-called {\it little group}, which is a subgroup of the Lorentz group leaving
some fixed $d$-momentum $p^a={\bf k}^a$ satisfying Eq.(\ref{prvic}) unchanged.

  Let us make the choice of ${\bf k}^a = (r k^0,0,\cdots,0,k^0)$, with $ r =\pm 1, k^0 > 0$. 
The infinitesimal generators of the little group 
$\omega_{ab} M^{ab},\,(\omega_{ij}=-\omega_{ji})$ can be found by requiring that, when operating on the $d$-vector 
${\bf k}^a$,
give zero so that accordingly the corresponding group transformations leave the $d$-vector ${\bf k}^a$ 
unchanged
\begin{eqnarray}
e^{-i \frac{1}{2} \omega_{bc} M^{bc}}\;  {\bf k}^a &=& {\bf k}^a,\quad {\rm or \;\; equivalently }
\nonumber
\\
\omega_{bc} M^{bc}{\bf k}^a &=& 0.
\label{lgp}
\end{eqnarray}
We see that Eq.(\ref{lgp}) requires 
\begin{equation}
  0 = \pmatrix{0 & -\omega_{01} & \ldots & -\omega_{0d}\cr
	       \omega_{10} & 0 & \ldots & \omega_{1d}\cr
		 \vdots & \vdots & \ddots & \vdots\cr
		 \omega_{d0} & \omega_{d1} & \ldots &0}
		 \pmatrix{r k^0\cr 0\cr \vdots\cr k^0} = 
		 \pmatrix{\omega_{0d}\cr r\omega_{10}+\omega_{1d}\cr\vdots\cr\omega_{d0}} k^0
\end{equation}
and it follows
\begin{eqnarray}
\omega_{0d}&=& 0,
\nonumber
\\
\omega_{i0}+r\omega_{id}&=&0,\quad i = 1,2,3,5,\cdots d-1.
\label{lgp1}
\end{eqnarray}
All $\omega_{ab} M^{ab}$ with $\omega_{ab}$ subject to conditions (\ref{lgp1}) form the
Lie algebra of the little group. We choose the following basis of the little group
Lie algebra
\begin{eqnarray}
\Pi_i = M^{0i} + r M^{id},\quad i = 1,2,3,5,\cdots,d-1
\nonumber
\\
{\rm and \; all\;\;} M^{ij}, \quad i, j = 1,2,3,5,\cdots,d-1.
\label{glg1}
\end{eqnarray}
One finds
\begin{eqnarray}
[\Pi_i, \Pi_j] = 0, \quad [\Pi_i, M^{jk}] = i (\eta^{ij} \Pi_k - \eta^{ik} \Pi_j).
\label{glgc}
\end{eqnarray}
We are interested only in discrete representations of the Poincar\' e group. This means that
the states in the representation space can be labeled by the momentum and an additional
label for internal degrees of freedom, which can only have {\it discrete} values.

\vskip0.5cm
{\it Lemma \ref{littlegroup}.1:} For a discrete representation of the Poincar\' e group
operators $\Pi_i$ give zero.

\vskip0.2cm

{\it Proof:} We may arrange the representation space of the little group
to be eigenvectors of the commuting operators
$\Pi_i$ of the little group\footnote{We can diagonalize $\Pi_i$ since we are interested in
unitary representations.}
\begin{equation}
\Pi_i |\Phi_a\rangle = b^i{}_a |\Phi_a\rangle.
\label{pibia}
\end{equation}
Here $a$ stands for a set of quantum numbers. We make 
the rotation $e^{i\theta M^{ij}}$ on a state $|\Phi_a\rangle$ depending on the parameter 
$\theta$ and the operator $M^{ij}$, with
the particular choice of
$i$ and $j$. Taking into account Eq.(\ref{glgc}) we find
\begin{eqnarray}
\Pi_i e^{i\theta M^{ij}} |\Phi_a\rangle = ( b^i{ }_a cos \theta - b^j{ }_a sin \theta ) e^{i\theta M^{ij}} |\Phi_a\rangle,
\nonumber\\
\Pi_j e^{i\theta M^{ij}} |\Phi_a\rangle = ( b^i{ }_a sin \theta + b^j{ }_a cos \theta )e^{i\theta M^{ij}} |\Phi_a\rangle.
\label{discrete}
\end{eqnarray}
It is obvious that the states $e^{i \theta M^{ij}} |\Phi_a\rangle$ produce a continous set of eigenvalues
for $\Pi_1, \ldots, \Pi_{d-1}$, 
since $\theta$ is a continous parameter, which contradicts the discreteness of the label $a$.
The only exception occurs when
$b^i{ }_a cos \theta - b^j{ }_a sin \theta = b^i{ }_a \;{\rm and} \quad 
b^i{ }_a sin \theta + b^j{ }_a cos \theta = b^j{ }_a$ for  all $\theta$.
This is only possible  if $b^i{ }_a = b^j{ }_a = 0 $ for all $a$ and each $i,j$.
 Therefore
\begin{eqnarray}
\Pi_i |\Phi_a\rangle = 0.
\label{discrete1}
\end{eqnarray}

\hfill\Qed
\vskip0.5cm
We see that on the representation space of the little group with the choice 
${\bf k}^a =(rk^0,o,\cdots,0,k^0)$, on which $(L^{0i} + r L^{id})|\Phi_a\rangle= (x^0 {\bf k}^i-
x^i {\bf k}^0 + rx^i {\bf k}^d -r x^d {\bf k}^i) |\Phi_a\rangle =0$ the following holds
\begin{equation}
\Pi_i |\Phi_a\rangle = \Pi^{(int)}_i |\Phi_a\rangle = 0, \quad \Pi^{(int)}_i = (S^{0i} + r S^{id}).
\label{piint}
\end{equation}

It follows then that the only little group generators, which are not necessarily zero on the 
representation space, are $M^{ij},
i, j = 1, 2, \ldots, d-1$ and they form the Lie algebra of $SO(d-2)$.
We conclude that {\it the irreducible discrete representations of the Poincar\' e group in 
$d(=2n)$ dimensions for massless particles are determined by the irreducible representations
of the group $SO(d-2)$. We will therefore denote the former with the same symbol as the 
latter with an additional index $r$ for the energy sign:
$(l_{n-1}, l_{n-2}, \ldots, l_2, l_1; r)_q$.}

The group $SO(d-2)$
has $\frac{(d-2)(d-3)}{2}$ generators and according to the commutation relations of 
Eq.(\ref{poincarealgebra}), $(\frac{d}{2} - 1)$ commuting 
generators and also $(\frac{d}{2} - 1)$ quantum numbers which determine a state $|\Phi_a\rangle$.

  We now show the validity of the Eq.(\ref{meq}).
  
\vskip0.5cm
{\it Lemma \ref{littlegroup}.2:} On the representation space of an irreducible
massless representation of the Poincar\' e group $(l_{n-1},\ldots,l_1; r)_q$
the equation (\ref{meq}) holds with
\begin{equation}
  \alpha = -\rho r 2^{n-1} (n-1)! (l_{n-1}+q(n-2))\ldots(l_2+q)l_1.
  \label{alfa1}
\end{equation}

\vskip0.2cm

{\it Proof:} First, we prove the lemma on the representation space of the little group
for the choice $p^a = (r k^0, 0,\ldots,0,k^0), r=\pm 1, k^0 > 0$.
We begin with the cases $a=1,2,\ldots,d-1$ in Eq.(\ref{meq})
\begin{eqnarray}
  W^a/\rho|\Phi_a\rangle = (\varepsilon^{a a_1}{ }_{a_2 a_3\ldots a_{d-1}} p_b M^{a_1 a_2} M^{a_3 a_4}\ldots)|\Phi_a\rangle = \nonumber\\
=  k^0(r\varepsilon^{a0}{ }_{a_1 a_2\ldots a_{d-3}a_{d-2}} M^{a_1 a_2} M^{a_3 a_4}\ldots
     - \varepsilon^{ad}{ }_{a_1 a_2\ldots a_{d-3}a_{d-2}} M^{a_1 a_2} M^{a_3 a_4}\ldots)|\Phi_a\rangle = \nonumber \\
=   2 k^0 (n-1) (r\varepsilon^{a0}{ }_{d a_1 a_2\ldots a_{d-3}}M^{d a_1}M^{a_2 a_3}\ldots
     -\varepsilon^{ad}{ }_{0 a_1 a_2\ldots a_{d-3}}M^{0 a_1}M^{a_2 a_3}\ldots)|\Phi_a\rangle = \nonumber \\
=   2 k^0 (n-1) (\varepsilon^{0d}{ }_{a a_1 a_2\ldots a_{d-3}}(r M^{d a_1}-M^{0 a_1})M^{a_2 a_3}\ldots)|\Phi_a\rangle=\nonumber\\
=   -2 k^0 (n-1) {\balpha}^{ai} \Pi_i|\Phi_a\rangle
  \label{wijenic}
\end{eqnarray}
where
\begin{equation}
{\balpha}^{ai} = 
  \varepsilon^{0dai}{ }_{a_1 a_2 \cdots a_{d-4}} M^{a_1a_2} \cdots M^{a_{d-5}a_{d-4}},\quad
[\Pi^i, {\balpha}^{ai}] = 0.
\nonumber
\end{equation}
 Eq.(\ref{discrete1}) concludes the proof for $a=1,2,\ldots,d-1$.

We now turn to the case $a=0$
\begin{eqnarray}
  W^0/\rho|\Phi_a\rangle = (\varepsilon^{0b}{ }_{a_1 a_2\ldots a_{d-3}a_{d-2}} p_b M^{a_1 a_2} M^{a_3 a_4}\ldots)|\Phi_a\rangle = \nonumber\\
=   -k^0 (\varepsilon^{0d}{ }_{a_1 a_2\ldots a_{d-3}a_{d-2}} M^{a_1 a_2} M^{a_3 a_4}\ldots)|\Phi_a\rangle = 
  (-r)p^0 \tilde{\Gamma}_{d-2}|\Phi_a\rangle.
  \nonumber
\end{eqnarray}
Eq.(\ref{alfa1}) holds according
to the value of the handedness operator obtained
in  section \ref{representations} and applied to the group $SO(d-2)$.
The case $a=d$ goes similarly
\begin{eqnarray}
  W^d/\rho|\Phi_a\rangle = (\varepsilon^{db}{ }_{a_1 a_2\ldots a_{d-3}a_{d-2}} p_b M^{a_1 a_2} M^{a_3 a_4}\ldots)|\Phi_a\rangle = \nonumber\\
=   r k^0 (\varepsilon^{d0}{ }_{a_1 a_2\ldots a_{d-3}a_{d-2}} M^{a_1 a_2} M^{a_3 a_4}\ldots)|\Phi_a\rangle = 
  (-r)p^d \tilde{\Gamma}_{d-2}|\Phi_a\rangle.
  \nonumber
\end{eqnarray}

Again, we point out that in all these derivations $M^{ab}$ can be replaced by $S^{ab}$.

To prove Eq.(\ref{meq}) on the whole representation space, one only has to note that 
Eq.(\ref{meq}) is in covariant form and must therefore hold generally. To put this explicitly:
the linear hull of the states
$|\Phi_a'\rangle = U(\Lambda)|\Phi_a\rangle$, where $|\Phi_a\rangle$ runs through the representation space of
our little group and $\Lambda\in SO(1, d-1)$
(here $U(\Lambda)$ denotes the unitary transformation belonging to the
element $\Lambda\in SO(1,d-1)$ in our representation), is dense in the whole representation
space. It is therefore sufficient to prove (\ref{meq}) for $|\Phi_a'\rangle$.
This follows from
\begin{eqnarray}
  (W^a-\alpha p^a)|\Phi_a'\rangle = U(\Lambda)U(\Lambda)^{-1}(W^a-\alpha p^a)U(\Lambda)|\Phi_a\rangle=
  \nonumber\\
=  U(\Lambda) (\Lambda^{-1})_{ab}(W^b-\alpha p^b)|\Phi_a\rangle = 0. \nonumber
\end{eqnarray}

\hfill\Qed
\vskip0.5cm

  It is proved that Eq.(\ref{meq}) with the parameter $\alpha$ from Eq.(\ref{alfa1}) 
  holds on the Poincar\' e group representation
$(l_{n-1},\ldots,l_1; r)_q$. It is therefore a candidate for an equation of motion. 
Before it can be admitted to that status we have to investigate its solutions.
We do that in the next section.

\section{Equations of motion.} \label{equationsofmotions}

  In this section we investigate the solutions of the equations
\begin{equation}
  (W^a = \alpha p^a)|\Phi\rangle,\quad p_a p^a|\Phi\rangle = 0,
  \label{enacba}
\end{equation}
on the space with internal Lorentz group $SO(1, d-1)$ 
representation $(l_n, l_{n-1}, \ldots, l_2, l_1)_{+1}$.
Since Eq.(\ref{enacba}) is in a covariant form it is clear that the space
of its solutions forms a representation of the Poincar\' e group. It also corresponds to
massless particles as indicated by the second equation in Eq.(\ref{enacba}). What remains to
be investigated is the irreducibility and discreteness of solutions.

For irreducibility, we will allow only degeneracy in the sign of $p^0 $ (the energy) and no degeneracy
as far the internal degrees of freedom are concerned, since we don't want the same equation to
describe particles of different spins. As we will see, to provide both this and 
the discreteness, the condition
\begin{equation}
  \alpha = - \rho 2^{n-1} (n-1)! (l_{n-1}+(n-2))\ldots(l_2 + 1) l_1
  \label{alfa}
\end{equation}
  must be fulfilled. For $d=4$ this is also the sufficient condition, while for dimensions
$d\ge 6$ for sufficiency the following condition must be added
\begin{equation}
  l_1 \not= 0 \Longleftrightarrow \Gamma^{(int)}\not=0.  \label{pogoj}
\end{equation}
  As shown in  section \ref{littlegroup}, due to the covariant form of Eq.(\ref{enacba}) the 
investigation of the Poincar\' e representations formed by the solutions of Eq.(\ref{enacba}) reduces
to the investigation of the $SO(d-2)$-representations formed by the solutions of
Eq.(\ref{enacba}) with $p^a = {\bf k}^a = (r k^0, 0,\ldots, 0, k^0), r=\pm 1, k^0>0$.

  We begin with the following lemma.

\vskip0.5cm
{\it Lemma \ref{equationsofmotions}.1 :} On the space with internal Lorentz group $SO(1, d-1)$ 
representation $(l_n, l_{n-1}, \ldots, l_2, l_1)_r$,  states satisfying 
the discreteness condition of Eq.(\ref{piint}) for the little group, form the 
$SO(d-2)$-irreducible representation space $(l_{n-1}, \ldots, l_1)_r$, where the subgroup $SO(d-2)\le
SO(1,d-1)$ acts on coordinates $1, 2, \ldots, d-1$.

\vskip0.2cm

{\it Proof:}
  Eqs.(\ref{glgc}) imply that the space of solutions 
$\{|\Phi_a\rangle\}$ of Eq.(\ref{piint}) forms an $SO(d-2)$-invariant space, where 
$SO(d-2)\le SO(1, d-1)$ acts on coordinates $1, 2, \ldots, d-1$. To prove 
our lemma we must show that $\{|\Phi_a\rangle\}$ is irreducible and corresponds to the 
$SO(d-2)$-representation
$(l_{n-1}, \ldots, l_1)_r$. We do this by choosing any 
state $|\Phi_a\rangle\in\{|\Phi_a\rangle\}$ with a $SO(d-2)$-dominant weight (with $q=r$)
and proving both that it is unique
(up to a scalar multiple) and has $SO(d-2)$-weight $(l_{n-1}, \ldots, l_1)$. 
This is possible because the $SO(1, d-1)$-dominant weight state is
in $\{|\Phi_a\rangle\}$, as shown by Eqs.(\ref{dom2}) with $q=r$, 
Eq.(\ref{discrete1}) and Eq.(\ref{piint}). The set of states 
$\{|\Phi_a\rangle\}$ is therefore not trivial.
Since $|\Phi_a\rangle$ has a $SO(d-2)$-dominant weight, Eqs.(\ref{dom3}) and (\ref{dom4}) must hold
with $q=r$. Eq.(\ref{piint}) then implies Eq.(\ref{dom2}) for
$q=r$.
Therefore, the state $|\Phi_a\rangle$ has a $SO(1, d-1)$-dominant weight (with $q=r$)
and is therefore unique
(up to a scalar multiple). Thus we have also proven that the corresponding 
representation of the Poincar\' e group is determined by all but the first 
dominant weight component
of the internal Lorentz group representation. It is therefore
$(l_{n-1}, l_{n-2}, \ldots, l_2, l_1)_r$ as was to be shown.
\hfill\Qed
\vskip0.5cm

  Given any representation of the internal Lorentz group, say $(l_n, \ldots, l_1)_{+1}$, the
only possible discrete Poincar\' e representation are given by the previous lemma and
Eq.(\ref{equiv}): they are
\begin{equation}
(l_{n-1}, \ldots, l_1; +1)_{+1} \quad{\rm and}\quad
(-l_{n-1}, \ldots, -l_2, (-1)^{n-1} l_1; -1)_{-1}=(l_{n-1}, \ldots, l_2, -l_1; -1)_{+1}.
\label{resitve}
\end{equation}
We note that in four dimensions in our notation $(\sigma; r)_q$ of the Poincar\' e group representation
the quantum number $\sigma$ is helicity since it is the eigenvalue of the operator $S^{12}$. We have therefore
generalized the known result in four dimensions which states $\sigma = r l_1$ 
(\cite{weinberg2}, \cite{fonda}) to general even-dimensional spaces.

So, according to Eq.(\ref{alfa1}) if we are to hope for the solutions of Eq.(\ref{enacba}) 
to form a discrete Poincar\' e representation we must have
$  \alpha = -\rho r 2^{n-1} (n-1)! (l_n+n-1)\ldots(l_2+1)(r l_1)$, 
which is exactly Eq.(\ref{alfa}).

We now answer the question: when are the solutions of the equations
\begin{equation}
  (W^a = \alpha p^a)|\Phi\rangle, \quad \alpha = -\rho 2^{n-1} (n-1)! (l_{n-1}+(n-2))\ldots (l_2 + 1) l_1,
  \quad p_a p^a|\Phi\rangle = 0,
  \label{enacba2}
\end{equation}
on the space with internal Lorentz group $SO(1, d-1)$ 
representation $(l_n, l_{n-1}, \ldots, l_2, l_1)_{+1}$, 
exactly those described by Eq.(\ref{resitve}) ? First, we deal with the simplest case
of $d=4$.

\vskip0.5cm
{\em Lemma \ref{equationsofmotions}.2 :} On the space with internal Lorentz group $SO(1, 3)$ 
representation $(l_2, l_1)_{+1}$ the solutions of the Eq.(\ref{enacba2}) are exactly
those of Eq.(\ref{resitve}).

\vskip0.2cm

{\em Proof:} 
Again we choose
$p^a = {\bf k}^a = (r k^0, 0, \ldots, 0, k^0), \;r=\pm 1, \; k^0>0$.
In this case, the first equation of Eqs.(\ref{enacba2}), with $a=1,2$ reads as follows
\begin{eqnarray}
  (\Pi^{(int)}_2 &=& 0)|\Phi\rangle, \\
  (\Pi^{(int)}_1 &=& 0)|\Phi\rangle.
  \label{enacba3}
\end{eqnarray}
These are exactly  Eq.(\ref{piint}) and according to lemma \ref{equationsofmotions}.1 their solutions are
listed in Eq.(\ref{resitve}). Finally, we note that 
$(W^0 = \alpha p^0)|\Phi\rangle$ and $(W^3 = \alpha p^3)|\Phi\rangle$ both give 
 the equation $S^{12}|\Phi\rangle = r l_1 |\Phi\rangle$ which agrees with Eq.(\ref{resitve}) 
 and therefore doesn't impose any additional
constraints to the solutions of Eq.(\ref{resitve}). This concludes the proof.
\hfill\Qed
\vskip0.5cm

  We shall now prove  the general case $d\ge 6$.

\vskip0.5cm
{\it Lemma \ref{equationsofmotions}.3 :} On the space with the internal Lorentz group $SO(1, d-1)$
representation $(l_n, l_{n-1}, \ldots, l_1)_{+1}$, where
\begin{equation}
  l_1 \not= 0  \Longleftrightarrow \Gamma^{(int)} \not= 0, \nonumber
\end{equation}
the solutions of Eq.(\ref{enacba2}) are exactly those of 
Eq.(\ref{resitve}).

\vskip0.2cm

{\it Proof:} We prove the lemma for the standard choice of  the $d$-momentum
$p^a = (r k^0, 0, \ldots, 0, k^0),\; r=\pm 1,\; k^0>0$. From the proof of lemma \ref{littlegroup}.2
we know that
Eq.(\ref{enacba2}) reads as follows
\begin{equation}
  {\balpha}^{ij} \Pi^{(int)}_j |\Phi\rangle = 0, 
  \label{enacba4}
\end{equation}
\begin{equation}
  -\rho \varepsilon^{0d}{ }_{a_1 a_2\ldots a_{d-3}a_{d-2}} S^{a_1 a_2} S^{a_3 a_4}\ldots|\Phi\rangle = \alpha|\Phi\rangle,
  \label{enacba5}
\end{equation}
 with $ {\balpha}^{ij} 
  = \varepsilon^{0dij}{ }_{a_1 a_2\ldots a_{d-5}a_{d-4}}
    S^{a_1 a_2} S^{a_3 a_4} \ldots \quad{\rm for}\,\, i,j=1,2,\ldots,d-1.$
By lemma \ref{equationsofmotions}.1 it is sufficient to show that every solution $|\Phi\rangle$ of 
Eqs.(\ref{enacba4}, \ref{enacba5}) satisfies the condition
\begin{equation}
  \Pi^{(int)}_i|\Phi\rangle = 0, \quad {\rm for}\,\, i=1,2,\ldots,d-1.
  \label{enacba6}
\end{equation}
In what follows we take the definition of dominant weights  (Eq.(\ref{domweight}))
with $q=r$.
According to lemma \ref{equationsofmotions}.1 and the discussion directly following
this lemma the 
dominant weight of the  group $SO(1,d-1)$
satisfies both Eqs.(\ref{enacba4}, \ref{enacba5}), so that  we conclude that the space ${\cal W}$
of solutions
of Eq.(\ref{enacba4}) and Eq.(\ref{enacba5}) is nontrivial. It is also $SO(d-2)$-invariant since
the following relations hold
\begin{eqnarray}
 [{\balpha}^{ij} \Pi^{(int)}_j, S^{kl}] &=& i(\eta^{ik}{\balpha}^{lj} \Pi^{(int)}_j - \eta^{il}
{\balpha}^{kj} \Pi^{(int)}_j),
\nonumber  
\\
\;[ \varepsilon^{0d}{}_{a_1a_2a_3a_4 \ldots a_{d-3}a_{d-2}} S^{a_1a_2} S^{a_3a_4}\ldots S^{a_{d-3}a_{d-2}},
S^{kl} ] &=& 0,
\nonumber
\end{eqnarray}
for $i,j,k,l=1,2,\ldots,d-1$. To prove that Eq.(\ref{enacba6}) holds on $\cal W$
it suffices to prove that Eq.(\ref{enacba6}) holds on any $SO(d-2)$-dominant weight state
in $\cal W$. (By lemma \ref{equationsofmotions}.1 this also implies that $\cal W$
is irreducible and the $SO(d-2)$-dominant weight state in ${\cal W}$ is therefore unique up to a scalar multiple.)
We therefore choose any $SO(d-2)$-dominant weight state $|\Phi_l\rangle\in{\cal W}$. We designate
its weight as $(l_{n-1}', l_{n-2}',\ldots, l_2', l_1')_r$. It follows from Eq.(\ref{enacba5}) that
$ \alpha = -\rho 2^{n-1}(n-1)!(l_{n-1}'+r(n-2))\ldots(l_2'+r)l_1'$.
 
From Eq.(\ref{dom3}) it follows that
\begin{eqnarray}
  {\balpha}^{1i'}|\Phi_l\rangle &=& 2(n-2)\varepsilon^{0d}{ }_{1i'2j'\ldots}S^{2j'}\ldots|\Phi_l\rangle =
  \nonumber \\
   &=& 2r(n-2)\varepsilon^{0d}{ }_{1i'2j'\ldots} i S^{1j'}\ldots|\Phi_l\rangle 
    = -ir{\balpha}^{2i'}|\Phi_l\rangle ,
	\label{pogojdom2}
\end{eqnarray}
for $i'=3,5,\ldots,d-1$. If we add 
the equation with $i=2$ from Eq.(\ref{enacba4}), which we first  multiply by $ i r {\balpha}^{12}$, to 
the equation with $i=1$, we find taking into account Eq.(\ref{pogojdom2}) that
\begin{eqnarray}
  0 =  ((\Pi^{(int)}_2-ir\Pi^{(int)}_1){\balpha}^{12}
  + \Pi^{(int)}_{i'}({\balpha}^{1i'}+ir{\balpha}^{2i'}))|\Phi_l\rangle= 
  (\Pi^{(int)}_2-ir\Pi^{(int)}_1){\balpha}^{12}|\Phi_l\rangle.
 \label{prva}
\end{eqnarray}
Following similar techniques as in section \ref{representations}
and taking into account Eq.(\ref{enacba5}), we find
\begin{eqnarray}
  \alpha|\Phi_l\rangle &=&-\rho
  \varepsilon^{0d}{ }_{a_1a_2a_3a_4\ldots}S^{a_1a_2}S^{a_3a_4}\ldots|\Phi_l\rangle  = \nonumber \\
  &=& -\rho
  2(n-1)(S^{12}+r(n-2)) \varepsilon^{0d}{ }_{12a_1a_2\ldots}S^{a_1a_2}\ldots|\Phi_l\rangle  =
  \nonumber \\
  &=& -\rho 2(n-1)(l_{n-1}'+r(n-2)) {\balpha}^{12}|\Phi_l\rangle  \nonumber \\
  \Longrightarrow {\balpha}^{12}|\Phi_l\rangle &=& \frac{\alpha}{-\rho 2(n-1)(l_{n-1}'+r(n-2))}|\Phi_l\rangle. 
  \label{pogoj2}
\end{eqnarray}
Since the condition $l_1\not=0$ implies $\alpha\not=0$ (Eq.(\ref{alfa1})), we may conclude from Eq.(\ref{pogoj2}) and
Eq.(\ref{prva}) 
\begin{equation}
  (\Pi^{(int)}_2-ir\Pi^{(int)}_1)|\Phi_l\rangle = 0.  \label{glavna}
\end{equation}
From Eqs. (\ref{enacba4}), (\ref{pogojdom2}), (\ref{glavna}) we find
\begin{eqnarray}
  0 &=& (\Pi^{(int)}_1 {\balpha}^{1i'} + \Pi^{(int)}_2{\balpha}^{2i'} + 
  \Pi^{(int)}_{j'}{\balpha}^{j'i'})|\Phi_l\rangle  = \nonumber\\
    &=& (-ri\Pi^{(int)}_1 {\balpha}^{2i'} + \Pi^{(int)}_2{\balpha}^{2i'} + 
	\Pi^{(int)}_{j'}{\balpha}^{j'i'})|\Phi_l\rangle = \nonumber\\
    &=& -ir[\Pi^{(int)}_1, {\balpha}^{2i'}]|\Phi_l\rangle + {\balpha}^{2i'}(-ir\Pi^{(int)}_1 
	+ \Pi^{(int)}_2)|\Phi_l\rangle + 
    \Pi^{(int)}_{j'}{\balpha}^{j'i'}|\Phi_l\rangle  = \nonumber\\
    &=& -ir[\Pi^{(int)}_1, {\balpha}^{2i'}]|\Phi_l\rangle + \Pi^{(int)}_{j'}{\balpha}^{j'i'}
	|\Phi_l\rangle  = \nonumber\\
    &=& 2r(n-2)(\Pi^{(int)}_{j'}\varepsilon^{0d}{ }_{12i'j'a_1 a_2\ldots}S^{a_1 a_2}\ldots)|\Phi_l\rangle + 
      \Pi^{(int)}_{j'}{\balpha}^{j'i'}|\Phi_l\rangle =\nonumber\\
    &=& 2r(n-2)(\Pi^{(int)}_{j'}\varepsilon^{0d}{ }_{12i'j'a_1 a_2\ldots}S^{a_1 a_2}\ldots)|\Phi_l\rangle + 
	\nonumber \\
     && +  2r(n-2)(\Pi^{(int)}_{j'}(l_{n-1}'+r(n-3))\varepsilon^{0d}{ }_{12i'j'a_1 a_2\ldots}S^{a_1 a_2}\ldots)
	  |\Phi_l\rangle  = \nonumber\\
    &=& 2r(n-2)(l_{n-1}'+r(n-2))\Pi^{(int)}_{j'}(\varepsilon^{0d}{ }_{12i'j'a_1 a_2\ldots}S^{a_1 a_2}\ldots)
	|\Phi_l\rangle
    \nonumber
\end{eqnarray}
with indices $i', j' = 3, 5, \ldots, d-1$.

We summarize
\begin{eqnarray}
  \Pi^{(int)}_j(\varepsilon^{0d12ij}{ }_{a_1a_2a_3a_4\ldots}S^{a_1a_2}S^{a_3a_4}\ldots)|\Phi_l\rangle = 0  \quad
  {\rm for\,\,all\,\,} i.
  \label{ind1}
\end{eqnarray}
Now, we can repeat this process by using (\ref{dom4}) to obtain
\begin{eqnarray}
  \Pi^{(int)}_j(\varepsilon^{0d1234ij}{ }_{a_1a_2a_3a_4\ldots}S^{a_1a_2}S^{a_3a_4}\ldots)|\Phi_l\rangle = 0  \quad
  {\rm for\,\,all\,\,} i\label{ind2}
\end{eqnarray}
and so on. Finally we arrive at
\begin{eqnarray}
  \Pi^{(int)}_j(\varepsilon^{0d1234\ldots (d-8)(d-7)ij}{ }_{a_1a_2a_3a_4}S^{a_1a_2}S^{a_3a_4})|\Phi_l\rangle = 0  \quad
  {\rm for\,\,all\,\,} i,
  \label{ind3}  \\
  \Pi^{(int)}_j(\varepsilon^{0d1234\ldots (d-6)(d-5)ij}{ }_{a_1 a_2}S^{a_1 a_2})|\Phi_l\rangle = 0  \quad
  {\rm for\,\,all\,\,} i\label{ind4} 
\end{eqnarray}
and
\begin{eqnarray}
  \Pi^{(int)}_j(\varepsilon^{0d1234\ldots (d-4)(d-3)ij})|\Phi_l\rangle = 0  \quad
  {\rm for\,\,all\,\,} i.
\end{eqnarray}
This obviously implies
\begin{eqnarray}
   \Pi^{(int)}_{d-1}|\Phi_l\rangle = 0,\quad
    \Pi^{(int)}_{d-2}|\Phi_l\rangle = 0.\label{rezultat1}
\end{eqnarray}
We use this  result in Eq.(\ref{ind4}) for $i=d-4, d-3$ to find
\begin{equation}  
  \Pi^{(int)}_{d-3}S^{d-2\;d-1}|\Phi_l\rangle = 0,\quad
    \Pi^{(int)}_{d-4}S^{d-2\;d-1}|\Phi_l\rangle = 0. \label{xyz}
\end{equation}
Since $S^{d-2\;d-1}|\Phi_l\rangle = l_1'|\Phi_l\rangle$ and $\alpha\not=0$, Eq.(\ref{alfa1}) requires that
$l_1'\not=0$ and we conclude from Eq.(\ref{xyz})
\begin{equation}
  \Pi^{(int)}_{d-3}|\Phi_l\rangle = 0, \quad \Pi^{(int)}_{d-4}|\Phi_l\rangle = 0. \label{rezultat2}
\end{equation}
Using this result and the one of Eq.(\ref{rezultat1}) in Eq.(\ref{ind3}) for $i=d-6, d-5$, we find
\begin{eqnarray}
  \Pi^{(int)}_{d-5}\varepsilon^{0d1234\ldots(d-8)(d-7)(d-6)(d-5)}{ }_{a_1 a_2 a_3 a_4}S^{a_1 a_2}S^{a_3 a_4}|\Phi_l\rangle = 0,
  \nonumber  \\
  \Pi^{(int)}_{d-6}\varepsilon^{0d1234\ldots(d-8)(d-7)(d-6)(d-5)}{ }_{a_1 a_2 a_4 a_4}S^{a_1 a_2}S^{a_3 a_4}|\Phi_l\rangle = 0.
  \nonumber
\end{eqnarray}
Again, the equality  $\varepsilon^{0d1234\ldots(d-8)(d-7)(d-6)(d-5)}{ }_{a_1 a_2 a_3 a_4}S^{a_1 a_2}S^{a_3 a_4}|\Phi_l\rangle = 
 8(l_2'+r)l_1'|\Phi_l\rangle$ and $l_1'\not=0$ implies that
\begin{equation}
  \Pi^{(int)}_{d-5}|\Phi_l\rangle = 0, \quad \Pi^{(int)}_{d-6}|\Phi_l\rangle = 0 \label{rezultat3}
\end{equation}
and so on. Finally we find
\begin{equation}
  \Pi^{(int)}_1|\Phi_l\rangle = 0, \quad \Pi^{(int)}_2|\Phi_l\rangle = 0, \label{rezultat4}
\end{equation}
which proves $\Pi^{(int)}_i|\Phi_l\rangle=0$ for $i=1,2,\ldots,d-1$ and concludes the proof.
\hfill\Qed

\vskip0.3cm
We may now write down the main result of this paper:
\vspace{2mm}

{\it On the space with internal Lorentz group $SO(1, d-1)$
representation $(l_n, l_{n-1}, \ldots, l_1)_{+1},\,\,(l_1\not=0)$
the equation
\begin{equation}
  (W^a = \alpha p^a)|\Phi\rangle, \quad {\rm with} \quad \alpha = -\rho 2^{n-1}(n-1)!(l_{n-1}+n-2)\ldots(l_2+1)l_1
  \label{glavna_en2} 
\end{equation}
is the equation of motion for massless particles corresponding to the following representations of the
Poincar\' e group}
\begin{equation}
  (l_{n-1}, \ldots, l_1; +1)_{+1}\quad{\rm and}\quad (l_{n-1},\ldots,-l_1;-1)_{+1},
\end{equation}
{\it where the masslessness condition $(p_a p^a-0)|\Phi\rangle$ is not needed, since it follows from (\ref{glavna_en2}).

With the aid of Eqs. (\ref{gamaf}), (\ref{beta}) the equation of motion can also be written as
\begin{equation}
  (W^a = |\alpha| \Gamma^{(int)} p^a)|\Phi\rangle, \quad |\alpha| = \rho 2^{n-1}(n-1)!(l_{n-1}+n-2)\ldots(l_2+1)|l_1|.
  \label{glavna_en3} 
\end{equation}
This equation is convenient when dealing with positive and negative handedness on the same footing
(an example of this is the Dirac equation) since $|\alpha|$ is independent of the
sign of $l_1$.}

We note that the particular value of $\rho$ is irrelevant in Eqs. (\ref{glavna_en2}), (\ref{glavna_en3}) since $\rho$
is found in both the lefthand and the righthand side of the equations and thus cancels out. The value of $\rho$ becomes
relevant when dealing with the particular spin (i.e. in the next section) when it is used to insure that
the operators $S^i$  have the familiar values independent of the dimension.

Making a choice of $a=0$ one finds
\begin{equation}
(\overrightarrow{S}.\overrightarrow{p}= |\alpha| \Gamma^{(int)} p^0)|\Phi\rangle.
\label{sag}
\end{equation}
Since $(\Gamma^{(int)})^2=1$, one immediately finds that $|\alpha| = |\overrightarrow{S}.\overrightarrow{p}|/|p^0|$.
We shall comment that for spinors Eq.(\ref{sag}) is equivalent to Eq.(\ref{glavna_en3}), while for vectors 
Eq.(\ref{glavna_en3}) gives additional condition to Eq.(\ref{sag}).

\section{Discussions on examples.} \label{discussions}

We have shown in previous sections that massless fields in $d=2n$-dimensional spaces, if having the Poincar\' e symmetry,
obey the linear equations
of motion.
We shall present in this section the equations of motion for massless fields in $d=2n$ dimensional spaces for two 
types of fields, which we call\cite{mankoc92,mankoc93,mankoc94,mankoc95,mankoc99,mankocborstnik98} 
spinorial and vectorial fields, respectively.
Spinorial fields are defined by  the generators of the Lorentz transformations, which fulfil Eq.(\ref{spinor}) and 
are  represented by  $(\frac{1}{2},\ldots,\pm\frac{1}{2})_{+1}$.
The irreducible representations of the internal Lorentz group for vectorial fields  are given by 
$(1,\ldots,\pm 1)_{+1}$. We shall later  define the generators of the Lorentz group for the vectorial internal space
in any dimension $d=2n$
(see Eq.(\ref{vector1})).
 
Although both kinds of fields can be treated in an ordinary way, that is by using the group theoretical approaches,
which determine  properties of states of an irreducible representation by defining the operation of the 
 generators of the group 
on a state without going into any representation,  just as we have done
in sections above, we shall use the space of anicommuting $d=2n$
coordinates to describe the internal degrees of freedom (it is the spin degree of freedom in our case) for both 
types of fields mentioned above, representing states  as polynomials of the anticommuting coordinates.
We shall do that, because this way
offers a very simple and transparent presentation of operators in any dimension, as well as of representations. 
We shall follow the approach
of one of us\cite{mankoc92,mankoc93,mankoc99}, using  Grassmann coordinates.
As it was proven in ref.\cite{mankocnielsen99}
we could as well use differential forms instead of Grassmann coordinates. 
At the end we shall comment on results.

  We briefly review the description of the internal space we use \cite{mankoc99}. 
  It is the Grassmann space (also known as 
the exterior algebra)  spanned over $2n$-dimensional vector space; we denote it as $\Lambda_{2n}$.  
Formally, this space is the space spanned by $2n$ variables 
\begin{equation}
  \theta^0, \theta^1, \ldots, \theta^{2n}
\end{equation}
and their products, where the following anticommutation relations hold
\begin{equation}
  \lbrace \theta^a, \theta^b\rbrace = \theta^a \theta^b + \theta^a \theta^b = 0, \quad a, b=0, 1, \ldots, 2n. 
  \label{antikom}
\end{equation}
We use the symbol $\theta^a$ to also denote the operator of multiplication with the variable $\theta^a$. 
Accordingly, relations (\ref{antikom}) hold for $\theta^a$ as operators also. The operator of differentiation
$\frac{\partial}{\partial \theta^a}$  is defined as follows
\begin{equation}
\frac{\partial}{\partial\theta^a} (\theta^{a_1}\ldots\theta^{a_p}) = \left\lbrace\matrix{0, & {\rm if}\,\,
a\not= a_1, \ldots, a_k, \cr (-1)^{l-1} \theta^{a_1}\ldots\theta^{a_{l-1}}\theta^{a_{l+1}} \ldots\theta^{a_p}, 
& {\rm if}\,\, a = a_l}\right. ,
\end{equation}
if we assume that the differentiation is allways performed from the left.

We denote $p_{\theta a} = -i \frac{\partial}{\partial\theta^a}$. It is easily checked that the following
anticommutation relations hold
\begin{equation}
  \lbrace \theta^a, \theta^b\rbrace = \lbrace p_{\theta}^a, p_{\theta}^b\rbrace = 0, \quad
   \lbrace \theta^a, p_\theta^b \rbrace = -i \eta^{ab}.
\end{equation}

Following the references\cite{mankoc92,mankoc99}, we define  
the generators of the internal Lorentz group. For
spinors they are
\begin{equation}
  S_s^{ab} = -\frac{i}{4} [a^a, a^b], \quad {\rm where}\quad a^a = i(p_\theta^a - i\theta^a),
  \label{spinorfer}
\end{equation}
with $[a^a,a^b] = a^a a^b - a^b a^a$,
and for vectors they are
\begin{equation}
  S_v^{ab} = \theta^a p_\theta^b-\theta^b p_\theta^a.
  \label{vector1}
\end{equation}
Both $S_s^{ab}$ and $S_v^{ab}$  satisfy Lorentz group commutation relations and therefore furnish a representation
of the Lorentz group. For operators $a^a$ the following anticommutation relations hold
\begin{equation}
  \lbrace a^a, a^b\rbrace = 2\eta^{ab}.
\end{equation}

We shall first comment on the representations for the spinorial case and derive the corresponding equations of
motion for spinorial massless fields in $d=2n$. Later we shall do the same for vectorial massless fields.

\subsection{Spinors.} \label{Fermions}

  We introduce the following definitions for the handedness operator and the Pauli-Ljubanski vector
($\beta = i\frac{2^n}{d!}, \rho = \frac{2^{n-2}}{(d-2)!}$)
\begin{eqnarray}
  \Gamma^{(int)} = i\frac{2^n}{d!}\varepsilon_{a_1 a_2\ldots a_{d-1}a_d} S_s^{a_1 a_2} S_s^{a_3 a_4}\ldots 
  = -(-i)^{n+1} a^0 a^1\ldots a^{2n}, \\
  W^a = \frac{2^{n-2}}{(d-2)!} \varepsilon^{a a_1}{}_{a_2 a_3\ldots a_{d-1}} p_{a_1} S_s^{a_2 a_3} 
  S_s^{a_4 a_5}\ldots = \frac{(-i)^{n-1}}{2}p_b
   a^a a^b\,\,(
   a^0 a^1\ldots a^{2n}).
\end{eqnarray}
This ensures ${\Gamma^{(int)}}^2 = 1$ which implies $l_1\not=0$ for all
irreducible representations $(l_n, \ldots, l_1)_{+1}$ into which the representation space decomposes. Then we must have
$|l_1|\ge \frac{1}{2}$ and by the equation (\ref{gamaf}) this implies $|\Gamma^{(int)}|\ge 1$ where equality is achieved
for $l_n=l_{n-1}=\ldots=|l_1|=\frac{1}{2}$. Since this is our case we conclude that we are dealing with the 
representations
$(\frac{1}{2}, \ldots, \frac{1}{2}, \pm\frac{1}{2})_{+1}$. For $d=6$ we present in Table \ref{prva_tabela} the basis 
for two of the 
irreducible representations $(\frac{1}{2}, \frac{1}{2}, \pm\frac{1}{2})$ our internal space decomposes into.

  We now proceed to determination of the equations of motion. First we introduce the operators
\begin{equation}
\bar{S_s}^{ab} =  \frac{2^{n-2}}{(d-2)!} \varepsilon^{ab}{}_{a_1 a_2\ldots a_{d-2}} 
S_s^{a_1 a_2} S_s^{a_3 a_4} \ldots = 
\frac{(-i)^{n-1}}{2} a^a a^b\,\,
   (a^0 a^1\ldots a^{2n}).
\end{equation}
and state

\vskip0.5cm
{\em Lemma \ref{Fermions}.1 :} It holds
\begin{equation}
  \bar{S_s}^{ab} = -i\Gamma^{(int)} S_s^{ab}.
\end{equation}

\vskip0.2cm

{\em Proof:} Obvious from our previous definitions. \hfill \Qed
\vskip0.5cm

  We now state our equation of motion. It is Eq.(\ref{glavna_en3}) and it reads
\begin{equation}
  (W^a = p_b \bar{S_s}^{ab} = \frac{1}{2}\Gamma^{(int)} p^a)|\Phi\rangle.
\end{equation}
  Now, the lemma \ref{Fermions}.1 implies that this is equivalent to
\begin{equation}
  (-i p_b \Gamma^{(int)}S_s^{ab} = \frac{1}{2}\Gamma^{(int)} p^a)|\Phi\rangle.
\end{equation}
  Taking into account the definition $S_s^{ab} = -\frac{i}{2} (a^a a^b - \eta^{ab})$ and multiplying the 
above equation by $2\Gamma^{(int)}$ we obtain 
\begin{equation}
  (-p_b (a^a a^b - \eta^{ab}) = p^a)|\Phi\rangle
\end{equation}
which upon multiplication by $a_a$ implies
\begin{equation}
  p_b a^b |\Phi\rangle = 0. 
  \label{dirac}  
\end{equation}

By reversing this process we may show that Eq. (\ref{dirac}) implies Eq. (\ref{glavna_en3}). It is therefore
our equation of motion.  For $d=4$ we may recognize it as the Weyl-like equation.
Further discussions about this equation of motion can be found in references\cite{mankoc99,mankocnielsen99} 
and other references included in these papers. We may also check, that by multiplying Eq.(\ref{dirac}) 
by $-(-i)^{n+1} \prod_{a\not= 0} a^a$,  Eq.(\ref{dirac}) is equivalent to
\begin{equation}
(\Gamma^{(int)} p^0 = 2\overrightarrow{S_s}.\overrightarrow{p})|\Phi_\rangle.
\label{Weyl}
\end{equation}

Defining $\gamma^a$ matrices for any\cite{giorgi,mankocnielsen99} $d$ 
and replacing $a^a$ by $\gamma^a$,
we could repeat all the above derivations and end up with the Dirac equation for spinors in $d$-dimensional spaces.

\subsection{Vectors.} \label{Bosons}

  We take the following definition of the handedness operator ($\beta = \frac{i}{2^n n!^2}$)
\begin{equation}
  \Gamma^{(int)} = \frac{i}{2^n n!^2} \varepsilon_{a_1 a_2 \ldots a_{d-1}a_d} S_v^{a_1 a_2} S_v^{a_3 a_4}\ldots = 
 \frac{i}{n!^2}\varepsilon_{a_1 a_2 \ldots a_{d-1}a_d}
   \theta^{a_1} p_\theta^{a_2} \theta^{a_3} p_\theta^{a_4}\ldots .
\end{equation}
We limit our attention to the subspace $\Lambda_n$ spanned by $n$-monomials of the form
\begin{equation}
  \theta^{a_1} \theta^{a_2} \ldots \theta^{a_n}, \quad a_1, \ldots, a_n =  0, 1, \ldots, 2n. 
\end{equation}
On this subspace it obviously holds
\begin{equation}
  ((\Gamma^{(int)})^2 = 1)|\Phi\rangle,
  \label{gama_vektor} 
\end{equation}
from where we conclude that $l_1\not=0$ for all irreducible representations $(l_n, \ldots, l_1)_{+1}$ into which
$\Lambda_n$ decomposes. It is easily checked that
\begin{equation}
  (S_v^{ab} S_{v\,ab})^2 = S_v^{ab} S_{v\,ab}, \quad ({\rm no\,\,summation \,\,over\,\,} a, b)
  \label{blabla}
\end{equation}
which implies that (in absolute value) the possible values for $S_v^{ab}$ are $0, 1$.
Since $l_1$ is nonzero, we conclude from Eq.(\ref{gamaf}) that $|\Gamma^{(int)}|\ge 1$ where equality is achieved for
\begin{equation}
  l_n = l_{n-1} = \ldots = |l_1| = 1.
  \label{lji} 
\end{equation}
Therefore, equation (\ref{gama_vektor}) implies Eq.(\ref{lji}).
For $d=6$ we present in Table \ref{druga_tabela} the basis for the irreducible representation $(1, 1, 1)$.

  We now proceed to the formulation of the equation of motion on $\Lambda_n$. 
  We take $\rho = \frac{1}{2^{n-1}(n-1)!^2}$
in the definition of the Pauli-Ljubanski vector
\begin{eqnarray}
  W^a = \frac{1}{2^{n-1}(n-1)!^2} \varepsilon^{a a_1}_{a_2 a_3\ldots a_{d-1}} 
  p_{a_1} S_v^{a_2 a_3} S_v^{a_4 a_5}\ldots = \nonumber \\
  = \frac{1}{(n-1)!^2} \varepsilon^{a a_1}{}_{a_2 a_3\ldots a_{d-1}} p_{a_1} 
	\theta^{a_2} p_\theta^{a_3}
   \theta^{a_4} p_\theta^{a_5}\ldots = p_b
	\bar{S_v}^{ab}
\end{eqnarray}
where we have introduced
\begin{equation}
  \bar{S_v}^{ab} = \frac{1}{(n-1)!^2} \varepsilon^{ab}{}_{a_1 a_2\ldots a_{d-3}a_{d-2}} \theta^{a_1} 
  p_\theta^{a_2} \theta^{a_3} p_\theta^{a_4}\ldots .
\end{equation}
Before formulating the equations of motion we prove

\vskip0.5cm
{\em Lemma \ref{Bosons}.1 :} On the space $\Lambda_n$ it holds
\begin{equation}
  (\bar{S_v}^{ab} = -i\Gamma^{(int)} S_v^{ab})|\Phi\rangle.
  \label{kvak}
\end{equation}

\vskip0.2cm

{\em Proof:} We fix $a,b = 0, 1, \ldots, 2n$ and write
\begin{eqnarray}
  \Gamma^{(int)} = \frac{i}{n!^2} \left( n(n-1) p_\theta^a p_\theta^b\varepsilon_{ab a_1 a_2\ldots a_{d-2}} 
  \theta^{a_1}
\theta^{a_2} \theta^{a_3} p_\theta^{a_4}\ldots + \right.
  \nonumber \\
  \left.
  + n(n-1) \theta^a \theta^b\varepsilon_{ab a_1 a_2\ldots a_{d-2}} p_\theta^{a_1} p_\theta^{a_2} \theta^{a_3} 
  p_\theta^{a_4}\ldots + 
  n^2 S_v^{ab} \lbrack \bar{S}_{v\,ab} (n-1)!^2 \rbrack \right).
\end{eqnarray}
  Multiplying this by $S_{v\,ab}=\theta_a p_{\theta b} - \theta_b p_{\theta a}$ and $\bar{S_v\, ab}$ 
  respectively we
obtain
\begin{equation}
  \Gamma^{(int)} S_{v\,ab} = i(S_v^{ab} S_{v\,ab}) \bar{S}_{v\,ab}, \quad ({\rm no\,\,summation \,\,over\,\,} a, b)
  \label{prvaa}
\end{equation}
\begin{equation}
  \Gamma^{(int)} \bar{S}_{v\,ab} = i(\bar{S_v}^{ab} \bar{S}_{v\,ab}) S_{v\,ab}.
  \quad ({\rm no\,\,summation \,\,over\,\,} a, b)
  \label{drugaa}
\end{equation}
Taking into account the equations (\ref{gama_vektor}), (\ref{blabla}), (\ref{prvaa}), (\ref{drugaa}) we obtain
\begin{eqnarray}
  i \bar{S}_{v\,ab}|\Phi\rangle &=& (i \bar{S}_{v\,ab} (1-S_v^{ab}S_{v\,ab}) + 
  i \bar{S}_{v\,ab} S_v^{ab} S_{v\,ab})|\Phi\rangle = \nonumber \\
  &=& (-\Gamma^{(int)} \bar{S}_{v\,ab}\bar{S_v}^{ab} S_{v\,ab}(1-S_v^{ab}S_{v\,ab}) 
  + \Gamma^{(int)} S_{v\,ab})|\Phi\rangle = \nonumber \\
  &=& (-i \bar{S}_{v\,ab}\bar{S_v}^{ab}\bar{S}_{v\,ab}(S_v^{ab}S_{v\,ab})(1-S_v^{ab}S_{v\,ab}) + 
  \Gamma^{(int)} S_{v\,ab})|\Phi\rangle = 
  \nonumber\\
  &=& \Gamma^{(int)} S_{v\,ab}|\Phi\rangle,  
\quad ({\rm no\,\,summation \,\,over\,\,} a, b).
  \nonumber
\end{eqnarray}
\hfill \Qed
\vskip0.5cm

We may now write our equations of motion (\ref{glavna_en3}) 
\begin{equation}
  (p_b \bar{S_v}^{ab} = \Gamma^{(int)} p^a)|\Phi\rangle.
\end{equation}
With the equation (\ref{kvak}) this is equivalent to
\begin{equation}
  (p_b S_v^{ab} = i p^a)|\Phi\rangle
  \label{izhodna}
\end{equation}
or
\begin{equation}
  (p_b (\theta^a p_\theta^b - \theta^b p_\theta^a) = i p^a)|\Phi\rangle.
  \nonumber
\end{equation}
We multiply the last equation by $p_\theta^a \theta_a$ (no summation over $a$) to obtain
\begin{equation}
  (p_\theta^a (p_b \theta^b) = 0)|\Phi\rangle. 
  \label{vmesna_en} 
\end{equation}
Since this holds for every $a$ we may conclude
\begin{equation}
  (p_b \theta^b = 0)|\Phi\rangle. 
  \label{koncno1}
\end{equation}
We similarly obtain 
\begin{equation}
  (p_b p_\theta^b = 0)|\Phi\rangle. 
  \label{koncno2}
\end{equation}
The equations (\ref{koncno1}), (\ref{koncno2}) are also easily shown to imply (\ref{izhodna}) :
 one simply multiplies Eqs. (\ref{koncno1}), (\ref{koncno2}) by $p_{\theta a}, \theta_a$, respectively, and adds
 the two equations. Therefore (\ref{koncno1}) and (\ref{koncno2}) are the equations of motion.

  The above discussion also shows that Eq.(\ref{sag}),
  which is a special case $a=0$ of Eq.(\ref{glavna_en3}), is not
equivalent to Eq.(\ref{glavna_en3}) for vectors.
It is equivalent to (\ref{vmesna_en}) with $b=0$ which in general does not imply (\ref{koncno1}). 
Therefore in the case of vectors (and in general) Eq.(\ref{sag}) must be complemented by additional equations to 
ensure that it is the equation of motion. These additional equations are of course the remaining equations 
$a=1, 2, \ldots$ of (\ref{glavna_en3}).

  Finally we show that the obtained equations of motion are the generalized Maxwell equations
\cite{weinberg2}. The space $\Lambda_n$ may be
identified with the space of totally antisymmetric $n$-tensor $F^{a_1 a_2\ldots a_n}$ with the correspondence
\cite{matematika}
\begin{equation}
  \theta_{a_1} \theta_{a_2}\ldots\theta_{a_n}\in\Lambda_n \longleftrightarrow
  F^{c_1 c_2 \ldots c_n} = \frac{1}{n!}\varepsilon^{c_1 c_2\ldots c_n}{}_{b_1 b_2 \ldots b_n}
  \varepsilon^{b_1 b_2 \ldots b_n}{}_{a_1 a_2 \ldots a_n}.
  \nonumber
\end{equation}

  Then Eqs.(\ref{koncno1}), (\ref{koncno2}) may be written as 
\begin{equation}
  p_a F^{a a_1\ldots a_{n-1}} = p_a \varepsilon^{a a_1\ldots a_{n-1}}{}_{b_1 b_2\ldots b_n} 
  F^{b_1 b_2\ldots b_n} = 0,
\end{equation}
which are the generalized Maxwell equations.

\section{Acknowledgement.} This work was supported by Ministry of
Science and Technology of Slovenia. One of us (N.S.M.B.) wants to thank for the many 
discussions to Holger Bech Nielsen and Anamarija Bor\v stnik.

\begin{table}
\caption{Two  of the spinorial irreducible representations of the Lorentz group with $(\frac{1}{2},\frac{1}{2},
\frac{\pm 1}{2}) $ in the Grassmann space for $d=6$. 
The dominant weight
state is listed first, so the first representations corresponds to $(\frac{1}{2}, \frac{1}{2}, \frac{1}{2})$ and the
second to $(\frac{1}{2}, \frac{1}{2}, -\frac{1}{2})$.}
\begin{tabular}{ccccc}
\label{prva_tabela}
$|\Phi\rangle$ & $S_s^{06}$ & $S_s^{12}$ & $S_s^{35}$ & $\Gamma^{(int)}$\\
\tableline
   $(\theta^0 + \theta^6)(\theta^1 - i\theta^2)(\theta^3 - i\theta^5)$   & 
   $\frac{i}{2}$ & $\frac{1}{2}$ & $\frac{1}{2}$ & $-1$\\
   $(1 + \theta^0\theta^6)(\theta^1 - i\theta^2)(1 + i\theta^3\theta^5)$   
   & $-\frac{i}{2}$ & $\frac{1}{2}$ & $-\frac{1}{2}$ & $-1$\\
   $(\theta^0 + \theta^6)(1 + i\theta^1\theta^2)(1 + i\theta^3\theta^5)$   
   & $\frac{i}{2}$ & $-\frac{1}{2}$ & $-\frac{1}{2}$ & $-1$\\
   $(1 + \theta^0\theta^6)(1 + i \theta^1\theta^2)(\theta^3 - i\theta^5)$   
   & $-\frac{i}{2}$ & $-\frac{1}{2}$ & $\frac{1}{2}$ & $-1$\\
\tableline
   $(\theta^0 + \theta^6)(\theta^1 - i\theta^2)(\theta^3 + i\theta^5)$   
   & $\frac{i}{2}$ & $\frac{1}{2}$ & $-\frac{1}{2}$ & $1$\\
   $(1 + \theta^0\theta^6)(\theta^1 - i\theta^2)(1 - i\theta^3\theta^5)$   
   & $-\frac{i}{2}$ & $\frac{1}{2}$ & $\frac{1}{2}$ & $1$\\
   $(\theta^0 + \theta^6)(1 + i\theta^1\theta^2)(1 - i\theta^3\theta^5)$   
   & $\frac{i}{2}$ & $-\frac{1}{2}$ & $\frac{1}{2}$ & $1$\\
   $(1 + \theta^0\theta^6)(1 + i \theta^1\theta^2)(\theta^3 + i\theta^5)$  
   & $-\frac{i}{2}$ & $-\frac{1}{2}$ & $-\frac{1}{2}$ & $1$\\
\end{tabular}
\end{table}

\begin{table}
\caption{One of the vectorial  irreducible representations with $(1,1,1)$ of the Lorentz group in the Grassmann space 
for $d=6$. The dominant weight state is listed first.}
\begin{tabular}{ccccc}
\label{druga_tabela}
 $|\Phi\rangle$ & $S_v^{06}$ & $S_v^{12}$ & $S_v^{35}$ & $\Gamma^{(int)}$\\
\tableline
   $(\theta^0 + \theta^6)(\theta^1 - i\theta^2)(\theta^3 - i\theta^5)$   & $i$ & $1$ & $1$ & $-1$\\
   $(\theta^0 + \theta^6)(\theta^1 + i\theta^2)(\theta^3 + i\theta^5)$   & $i$ & $-1$ & $-1$ & $-1$\\
   $(\theta^0 - \theta^6)(\theta^1 - i\theta^2)(\theta^3 + i\theta^5)$   & $-i$ & $1$ & $-1$ & $-1$\\
   $(\theta^0 - \theta^6)(\theta^1 + i\theta^2)(\theta^3 - i\theta^5)$   & $-i$ & $-1$ & $1$ & $-1$\\
   $(\theta^0\theta^6 + i\theta^1\theta^2)(\theta^3 - i\theta^5)$   & $0$ & $0$ & $1$ & $-1$\\
   $(\theta^0\theta^6 + i\theta^3\theta^5)(\theta^1 - i\theta^2)$   & $0$ & $1$ & $0$ & $-1$\\
   $(\theta^0 + \theta^6)(\theta^1\theta^2 - \theta^3\theta^5)$   & $i$ & $0$ & $0$ & $-1$\\
   $(\theta^0\theta^6 - i\theta^1\theta^2)(\theta^3 + i\theta^5)$   & $0$ & $0$ & $1$ & $-1$\\
   $(\theta^0\theta^6 - i\theta^3\theta^5)(\theta^1 + i\theta^2)$   & $0$ & $1$ & $0$ & $-1$\\
   $(\theta^0 - \theta^6)(\theta^1\theta^2 + \theta^3\theta^5)$   & $i$ & $0$ & $0$ & $-1$\\
\end{tabular}
\end{table}

\end{document}